\documentclass[journal]{IEEEtran}

\usepackage{fancyhdr}
\usepackage{amsmath}
\usepackage{amssymb}
\usepackage{graphicx}
\usepackage{booktabs}
\usepackage{amsfonts}
\usepackage{multirow}
\usepackage{algorithm}
\usepackage{algorithmic}
\usepackage{mathrsfs}
\usepackage{bm}
\usepackage{exscale}
\usepackage{relsize}
\usepackage{setspace}
\usepackage{color}
\usepackage{ulem}
\newtheorem{lemma}{\textbf{Lemma}}

\newtheorem{proposition}{\textbf{Proposition}}

\ifCLASSINFOpdf
\else
\fi

\begin{document}

\title{On the Achievable Spectral Efficiency of Spatial Modulation Aided Downlink Non-Orthogonal Multiple Access}

\author{
Xuesi~Wang,~\IEEEmembership{Student~Member,~IEEE},
Jintao~Wang,~\IEEEmembership{Senior~Member,~IEEE},
Longzhuang~He,~\IEEEmembership{Student~Member,~IEEE},
Zihan~Tang,~\IEEEmembership{Student~Member,~IEEE},
and Jian~Song,~\IEEEmembership{Fellow,~IEEE}

\thanks{
X. Wang, J. Wang, L. He, Z. Tang and J. Song are with the Department of Electronic Engineering, Tsinghua University, Beijing, 100084, China (e-mail: wxs.tsinghua@gmail.com).

This work was supported by the National Natural Science Foundation of China (Grant No. 61471221) and the Tsinghua Fudaoyuan Research Fund.
}
}

\maketitle
\begin{abstract}
In this paper, a novel spatial modulation aided non-orthogonal multiple access (SM-NOMA) system is proposed. We use mutual information (MI) to characterize the achievable spectral efficiency (SE) of the proposed SM-NOMA system. Due to the finite-alphabet space-domain inputs employed by SM, the expression of the corresponding MI lacks a closed-form formulation. Hence, a lower bound is proposed to quantify the MI of the SM-NOMA system. Furthermore, its asymptotic property is also theoretically investigated in both low and high signal-to-noise ratio (SNR) regions. The SE performance and its analysis of our proposed SM-NOMA system are confirmed by simulation results.
\end{abstract}

\IEEEpeerreviewmaketitle

\section{Introduction}
Non-orthogonal multiple access (NOMA) constitutes a promising technique in the fifth-generation (5G) mobile networks~\cite{Ding2017NOMAMagazine}. Different from the traditional orthogonal multiple access (OMA), different users are designed to access the same time, frequency and code domain resources, but different power levels in NOMA. Compared to OMA, NOMA is more flexible and is capable of offering a higher sum rate and lower outage probability~\cite{Ding2017NOMAMagazine}\cite{Ding2016UserPairing}.

Moreover, the combination of NOMA and multiple-input multiple-output (MIMO) has recently attracted substantial research interest~\cite{Chen2016QuasiDegradation}. However, in conventional MIMO systems, the simultaneous use of multiple transmit antennas (TAs) requires a large amount of radio frequency (RF) chains, which significantly increases the corresponding power dissipation and the implementation complexity. To circumvent this problem, spatial modulation (SM) technique has recently been applied to MIMO systems~\cite{Renzo2011SMSurvey}-\cite{Yang2015DGforSM}. In conventional SM, only one TA is activated for each symbol's transmission, hence, only one single RF chain is needed. Since the active antenna is uniform-randomly selected from the transmit antennas in an SM transmitter, the information is thus carried both by the active antenna indices and by the transmitted amplitude-phase modulation (APM) symbols.

For SM-MIMO systems, lots of literature has considered the spectral efficiency (SE) of SM systems through mutual information (MI)~\cite{He2016TWC}\cite{Basnayaka2016FewActive}. In~\cite{Basnayaka2016FewActive}, the authors explored the channel capacity of SM associated with a large array of antennas and maximized the mutual information by optimizing the distribution of the channel input. In addition, the current research on multiple access methods based on SM mainly focuses on the uplink transmission~\cite{GR2015SMMAchannels} and the receive SM (R-SM) schemes over the MIMO broadcast channels~\cite{Stavridis2016RSM}. To the best of our knowledge, NOMA is a novel multiple access method for downlink multi-user SM systems.

In this paper, we propose a novel SM aided NOMA (SM-NOMA) system and use MI to characterize its achievable SE. As the exact expression of the corresponding MI lacks a closed-form formulation, computational-massive methods, such as numerical integration or Monte Carlo method, are usually required. In this context, a closed-form lower bound is proposed in this paper to provide an accurate approximation to the exact MI of the proposed system. Meanwhile, the asymptotic values in both high and low signal-to-noise ratio (SNR) regions of the MI are also derived to characterize the tightness of the proposed closed-form MI' lower bound with respect to its true value. Finally, the SE performance and its analysis of our proposed SM-NOMA system are validated by simulation results.

The rest of this paper is organized as follows. Section II describes the proposed SM-NOMA system model. Section III analyzes its achievable SE through MI. In Section IV, the simulation results are provided to validate the proposed expressions and the system's performance. Finally, Section V concludes this paper.

\small

\begin{figure*}[t]
    \begin{equation}
    \arraycolsep=1.0pt\def\arraystretch{2.5}
    \begin{array}{rcl}
	\begin{aligned}
      \displaystyle \Omega_{r,k} &\sim \left( \prod_{t=k+1}^{K} \frac{1}{N_t}\right) \cdot \sum_{n_{k+1}=1}^{N_{k+1}} \cdots \sum_{n_{K}=1}^{N_{K}} \mathcal{CN} \left(0, \sigma_{\text{v}}^2 + \sigma_{\text{s}}^2 \sum_{t=k+1}^{K} b_{r,t}^{(n_t)2} \alpha_t^2 \right),~~\Omega_{K,K} \sim \mathcal{CN} \left(0, \sigma_{\text{v}}^2 \right) \\
      \displaystyle Y_{r,k} &\sim \left( \prod_{t=k}^{K} \frac{1}{N_t} \right) \cdot \sum_{n_{k}=1}^{N_{k}} \cdots \sum_{n_{K}=1}^{N_{K}} \mathcal{CN} \left(0, \sigma_{\text{v}}^2 + \sigma_{\text{s}}^2 \sum_{t=k}^{K} b_{r,t}^{(n_t)2} \alpha_t^2 \right).
    \end{aligned}
	\end{array}
    \label{eq:Omega_Y}
    \end{equation}
\hrulefill
\end{figure*}

\normalsize

\section{System Model}
In this paper, we consider a base station (BS) equipped with $M$ TAs. Since the NOMA system is interference-limited, we adopt the hybrid multiple access scheme~\cite{Ding2016UserPairing}, i.e., users in one cell are divided into several groups where NOMA is implemented within each group and the inter-group interference can be eliminated by adopting OMA among different groups. We assume that each group contains $K$ users, while each user is equipped with a single antenna.

In our proposed SM-NOMA system, the signal vector transmitted by the BS can be considered as the superposition of the SM-signal vectors intended for the multiple users. Mathematically, the transmit signal vector can be expressed as follows:
\begin{equation}\label{eq:TS}
	\mathbf{x} = \sum_{k = 1}^K \alpha_k s_k \mathbf{w}_k^{(n_k)},
\end{equation}
where $s_k$ denotes the $k$-th user's transmitted APM symbol, $E\left[|s_k|^2\right] = 1$, and $\alpha_k$ denotes the $k$-th user's transmission power. Besides, $\mathbf{w}_k^{(n_k)} \in \mathbb{C}^{M \times 1}$ represents the space-domain input signal vector of the $k$-th user, which is uniform-randomly selected from $\Gamma_k \triangleq \{ \mathbf{w}_k^{(1)}, \mathbf{w}_k^{(2)}, \cdots, \mathbf{w}_k^{(N_k)} \}$. The nonzero elements subscripts of $\mathbf{w}_k^{(n_k)}$ represent the active antennas indices and we have normalized the energy, i.e., $\|\mathbf{w}_k^{(n)}\|_2^2 = 1$ ($n = 1, 2, \cdots, N_k$). In addition, $N_k$ represents the total amount of $\mathbf{w}_k^{(n_k)}$ belongs to the $k$-th user.

The design of $\Gamma_k$ is closely related to the specific space-domain alphabet design adopted by user $k$. In this paper, the conventional SM regime is adopted by the $k$-th user, i.e., we have $\Gamma_k = \{\mathbf{e}_1, \mathbf{e}_2, \ldots, \mathbf{e}_M\}$, where $\mathbf{e}_n$ represents the $n$-th column of an $(M \times M)$-dimensional identity matrix. In this case, each user randomly activates one of the $M$ TAs. Moreover, the number of required RF chains can be represents as $N_{\text{RF}} = \|\mathbf{x}\|_0$. In our adopted conventional SM regime, obviously, the number of RF chains used is equal to the number of users currently served, i.e., $N_{\text{RF}} = K$.

Upon assuming a flat-fading MIMO channel, the $k$-th user's received symbol can be represented as
\begin{equation}\label{eq:RS_k}
	y_k = \mathbf{h}_k^T \mathbf{w}_k^{(n_k)} \alpha_k s_k + \underbrace{\sum_{t \neq k} \mathbf{h}_k^T \mathbf{w}_t^{(n_t)} \alpha_t s_t}_{\text{intra-group interference}} + v_k,
\end{equation}
in which $\mathbf{h}_k^T \in \mathbb{C}^{1 \times M}$ denotes the narrowband channel vector associated with the $k$-th users, $v_k$ denotes the independent identically distributed (i.i.d) additive white Gaussian noise (AWGN), and its corresponding random variables $V \sim \mathcal{CN}(0,\sigma_{\text{v}}^2)$.

The intra-group interference can be partly eliminated by employing SIC~\cite{Chen2016QuasiDegradation}. Without loss of generality, we assume the decoding order as $(1, \cdots, K)$. A user can thus successfully decode the messages intended for those users having a smaller decoding order than himself, while the messages intended for the remaining users are simply handled as interference~\cite{Chen2016QuasiDegradation}.

To realize this assumption, the message $s_k$ intended for the $k$-th user must be decoded by the $r$-th ($k \leq r \leq K$) user correctly as
\begin{equation}\label{eq:RS_k_SIC}
	y_{r,k} = \mathbf{h}_r^T \mathbf{w}_k^{(n_k)} \alpha_k s_k + \sum_{t = k+1}^K \mathbf{h}_r^T \mathbf{w}_t^{(n_t)} \alpha_t s_t + v_r.
\end{equation}
Let $b_{r,k}^{(n_k)} = \mathbf{h}_r^T \mathbf{w}_k^{(n_k)}$ and $\omega_{r,k} = \sum_{t = k+1}^K b_{r,t}^{(n_t)} \alpha_t s_t + v_r$ denotes the intra-group interference as well as the AWGN, we get $y_{r,k} = b_{r,k}^{(n_k)} \alpha_k s_k + \omega_{r,k}$.

According to \eqref{eq:RS_k_SIC}, the SE of the $r$-th user decoding the $k$-th message can be characterized via MI between the $k$-th message received by the $r$-th user, i.e., $y_{r,k}$, and the transmitted APM-domain message $s_k$ comes with the space-domain message $\mathbf{w}_k^{(n_k)}$~\cite{He2016TWC}\cite{Basnayaka2016FewActive}, which is represented by $I_{r, k}$ and can be given as follows:
\begin{equation}\label{eq:MI_rk}
	I_{r,k} = I \left( Y_{r,k}; X, B_{r,k} \right),
\end{equation}
in which $y_{r,k}$, $s_k$, and $b_{r,k}^{(n_k)}$ are realizations of the random variables $Y_{r,k}$, $X$, and $B_{r,k}$, respectively. Furthermore, the achievable SE performance of the proposed system is characterized by the sum MI of all the users, i.e., $\sum_{k=1}^{K} I_{k, k}$.

\section{Mutual Information Analysis}
Because $I \left( Y_{r,k}; X, B_{r,k} \right) = h \left( Y_{r,k} \right) - h \left( Y_{r,k}| X, B_{r,k} \right)$, \eqref{eq:MI_rk} can be simplified as
\begin{equation}\label{eq:MI_simplified}
	I \left( Y_{r,k}; X, B_{r,k} \right) = h \left( Y_{r,k} \right) - h \left( \Omega_{r,k} \right),
\end{equation}
where the random variable $\Omega_{r,k}$ corresponds to the random realization $\omega_{r,k}$. Therefore, the MI analysis can be decomposed into the entropy calculation of the receive signal $y_{r,k}$ as well as the interference $\omega_{r,k}$.

In SM, for each symbol's transmission to the $k$-th user, the space-domain information $\mathbf{w}_k^{(n_k)}$ randomly selects one of the $N_k$ antenna selection vectors in $\Gamma_k$, according to a uniform probability distribution~\cite{He2016TWC}. Hence, $\mathcal{P}\left(B_{r,k} = b_{r,k}^{(n_k)} \right) = \frac{1}{N_k}$. Besides, we assume a complex-valued Gaussian input, i.e., $X \sim \mathcal{CN}(0, \sigma_{\text{s}}^2)$. 

Meanwhile, when the value of $B_{r,t}$ is determined, $\Omega_{r,k}$ and $Y_{r,k}$ can be seen as the superposition of several complex Gaussian random variables, e.g., $\left(\Omega_{r,k}\left| B_{r,k+1}, \cdots, B_{r,K}\right) \right. \sim \mathcal{CN} \left(0, \sigma_{\text{v}}^2 + \sigma_{\text{s}}^2 \sum_{t=k+1}^{K} b_{r,t}^{(n_t)2} \alpha_t^2 \right)$. Therefore, $\Omega_{r,k}$ and $Y_{r,k}$ are subject to Gaussian mixture distribution (GMD)~\cite{Kim2015EntropyGMD}, as given in \eqref{eq:Omega_Y}.


Without loss of generality, the probability density function (PDF) of a scalar complex GMD $A$ can be represented as~\cite{Kim2015EntropyGMD}
\begin{equation}
	f_A(a) = \sum_{l=1}^{L} \beta_l f_l(a) = \sum_{l=1}^{L} \beta_l \frac{1}{\pi \sigma_l^2} e^{- \frac{|a-\mu_l|^2}{\sigma_l^2}},
\end{equation}
where $\sum_{l=1}^L \beta_l = 1$, and $f_l(a) = \frac{1}{\pi \sigma_l^2} e^{- \frac{|a-\mu_l|^2}{\sigma_l^2}}$.

The entropy of $A$ can be represented as $h(A) = - \int f_A(a) \log_2 \left[f_A(a)\right] \mathrm{d}a$. Due to the logarithm of a sum of exponential funcations, the entropy of a GMD random variable has no closed-form formulation~\cite{Kim2015EntropyGMD}. It thus relies on numerical integration or Monte Carlo method. In order to avoid the huge computational complexity required by Monte Carlo method, we use \textbf{Lemma} \ref{lemma1} in~\cite{Huber2008Entropy} to provide the lower and upper bounds of $h(A)$:

\begin{lemma}
	The lower and upper bounds of $h(A)$ are given by:
	\begin{equation}
	\begin{split}
		h_{\text{LB}}(A) &= - \sum_{l=1}^{L} \beta_l \log_2 \left( \sum_{t=1}^{L} \beta_t z_{lt} \right) \\
		h_{\text{UB}}(A) &= \sum_{l=1}^{L} \beta_l \log_2\left( \frac{\pi e \sigma_l^2} {\beta_l} \right),
    \end{split}
	\end{equation}
    where $z_{lt} = \int f_l(a) f_t(a) \mathrm{d}a$.
\label{lemma1}
\end{lemma}
\begin{IEEEproof}
	Because the function $\log_2 (\cdot)$ is concave, according to Jensen's inequality, we have a lower bound as
	\begin{equation}
	\begin{split}
		h(A) &\textstyle{= - \sum_{l=1}^{L} \beta_l \int f_l(a) \log_2 \left[f_A(a)\right] \mathrm{d}a} \\
			&\textstyle{\geq - \sum_{l=1}^{L} \beta_l \log_2 \left[ \int f_l(a) f_A(a) \mathrm{d}a \right]} \\
			&\textstyle{= - \sum_{l=1}^{L} \beta_l \log_2 \left( \sum_{t=1}^{L} \beta_t z_{lt} \right)}.
	\end{split}
	\end{equation}
	Similarly, we have an upper bound as
	\begin{equation}
	\begin{split}
		h(A) &\textstyle{= - \sum_{l=1}^{L} \beta_l \int f_l(a) \log_2 \left[f_A(a)\right] \mathrm{d}a} \\
			&\textstyle{\leq - \sum_{l=1}^{L} \beta_l \int f_l(a) \log_2 \left[ \beta_l f_l(a) \right] \mathrm{d}a} \\
			&\textstyle{= \sum_{l=1}^{L} \beta_l \log_2\left( \frac{\pi e \sigma_l^2} {\beta_l} \right)}.
	\end{split}
	\end{equation}
\end{IEEEproof}

Considering the distribution of $b_{r,k}^{(n_k)}$ and $s_k$, we can simplify \textbf{Lemma} \ref{lemma1} as Proposition \ref{proposition1} by setting $\beta_l = \frac{1}{L}$ and $\mu_l = 0$:

\begin{proposition}
	The lower and upper bounds of $h(A)$ can be simplified as
	\begin{equation}\label{eq:bound_sim}
	\begin{split}
		h_{\text{LB}}(A) &= \log_2\left(\pi L\right) - \frac{1}{L} \sum_{l=1}^L \log_2\left( \sum_{t=1}^L \frac{1} {\sigma_l^2+\sigma_t^2} \right) \\
		h_{\text{UB}}(A) &= \log_2\left(\pi e L\right) + \frac{1}{L} \sum_{l=1}^L \log_2\left( \sigma_l^2 \right).
    \end{split}
	\end{equation}
\label{proposition1}
\end{proposition}

According to \eqref{eq:MI_simplified} and based on \eqref{eq:bound_sim}, by substituting $\Omega_{r,k}$ and $Y_{r,k}$ into $I_{\text{LB}} \left( Y_{r,k}; X, B_{r,k} \right) = h_{\text{LB}} \left( Y_{r,k} \right) - h_{\text{UB}} \left( \Omega_{r,k} \right)$, we propose the lower bound of MI. Considering the user pairing strategies in NOMA\footnote{The impact of user pairing on the performance of SM-NOMA systems is beyond the scope of this paper, which will be handled with more consideration in our future work.}~\cite{Ding2016UserPairing}\cite{Chen2016QuasiDegradation}, we derive the mathematical expression of $I_{\text{LB}}$ when $K = 2$, as given in \eqref{eq:I_LB}, where $i \in \{ 1,2 \}$, $\rho = \frac{\sigma_{\text{s}}^2}{\sigma_{\text{v}}^2}$ denotes the SNR and $b_{r,k}^{(n_tm_t)2} = b_{r,k}^{(n_t)2} + b_{r,k}^{(m_t)2}$. It is worth noting that $h(\Omega_{2,2}) = \log_2(\pi e \sigma_v^2)$ because $\Omega_{2,2} \sim \mathcal{CN} \left(0, \sigma_{\text{v}}^2 \right)$~\cite{Kim2015EntropyGMD}.

Moreover, to better characterize the tightness of the proposed closed-form MI' lower bound with respect to its true value, by reducing $\rho$ to zero and increasing $\rho$ without limit in \eqref{eq:I_LB}, we arrive at Proposition \ref{proposition2} and Proposition \ref{proposition3}. 

\begin{proposition}
	In low SNR region, the asymptotic values of MI are all $0$ while their lower bounds approach:
	\begin{equation}
	\begin{split}
		I_{\text{LB}} \left( Y_{i,1}; X, B_{i,1} \right) &\approx 1 - \log_2(e N_2) \\
		I_{\text{LB}} \left( Y_{2,2}; X, B_{2,2} \right) &\approx 1 - \log_2(e).
    \end{split}
	\end{equation}
\label{proposition2}
\end{proposition}

\begin{proposition}
	In high SNR region, the asymptotic values of $I \left( Y_{2,2}; X, B_{2,2} \right)$ and $I_{\text{LB}} \left( Y_{2,2}; X, B_{2,2} \right)$ do not exist while others approach:
    \begin{equation}\label{eq:HighSNR}
	\begin{split}
		I \left( Y_{i,1}; X, B_{i,1} \right) &\approx \log_2(1 + \frac{\alpha_1^2}{\alpha_2^2}) \\
		I_{\text{LB}} \left( Y_{i,1}; X, B_{i,1} \right) &\approx \log_2(1 + \frac{\alpha_1^2}{\alpha_2^2}) + 1 - \log_2(e N_2).
    \end{split}
	\end{equation}
\label{proposition3}
\end{proposition}
\begin{IEEEproof}
	Assuming $\sigma_{\text{s}}^2 \gg \sigma_{\text{v}}^2$, we can omit $\sigma_{\text{v}}^2$ in \eqref{eq:Omega_Y}. Besides, based on our adopted conventional SM regime, we can obtain $\Omega_{i,1} \sim \mathcal{CN} \left(0, \sigma_{\text{s}}^2 b^2 \alpha_2^2 \right)$, and $Y_{i,1} \sim \mathcal{CN} \left(0, \sigma_{\text{s}}^2 b^2 (\alpha_1^2 + \alpha_2^2) \right)$ according to~\cite{Hershey2007KL}, where $b^2 = \frac{1}{N_1} \sum_{n_1=1}^{N_1}b_{i,1}^{(n_1)2} \approx \frac{1}{N_2} \sum_{n_2=1}^{N_2}b_{i,2}^{(n_2)2}$.
	Therefore,
	\begin{equation}
	\begin{split}
		I \left( Y_{i,1}; X, B_{i,1} \right) &= h \left( Y_{i,1} \right) - h \left( \Omega_{i,1} \right) \\
		&= \log_2\left[\pi e \sigma_{\text{s}}^2 b^2 (\alpha_1^2 + \alpha_2^2)\right] -\log_2\left(\pi e \sigma_{\text{s}}^2 b^2 \alpha_2^2\right) \\
		&= \log_2(1 + \frac{\alpha_1^2}{\alpha_2^2}). \nonumber
	\end{split}
	\end{equation}
	Similarly, based on \eqref{eq:I_LB}, we can thus obtain the approximation of $I_{\text{LB}} \left( Y_{i,1}; X, B_{i,1} \right)$.
\end{IEEEproof}

Furthermore, it is observed that a constant shift exists between the MI and its lower bound. Particularly, the constant shift can be written as $C_{i,1} = 1 - \log_2(e N_2)$ and $C_{2,2} = 1 - \log_2(e)$. More importantly, a constant shift imposes no impact on the optimization of the MI's lower bound.

\small

\begin{figure*}[t]
    \begin{equation}
    \arraycolsep=1.0pt\def\arraystretch{2.5}
    \begin{array}{rcl}
	\begin{aligned}
		\displaystyle I_{\text{LB}} \left( Y_{i,1}; X, B_{i,1} \right) &= \log_2\left( \frac{N_1}e \right) - \frac{1}{N_1N_2} \cdot \sum_{n_{1}=1}^{N_{1}} \sum_{n_{2}=1}^{N_{2}} \log_2 \left[ \sum_{m_{1}=1}^{N_{1}} \sum_{m_{2}=1}^{N_{2}} \frac{1 + \rho \alpha_2^2 b_{i,2}^{(n_2)2}} {2 + \rho \left(\alpha_1^2 b_{i,1}^{(n_1m_1)2} + \alpha_2^2 b_{i,2}^{(n_2m_2)2} \right)} \right] \\
		\displaystyle I_{\text{LB}} \left( Y_{2,2}; X, B_{2,2} \right) &= \log_2\left( \frac{N_2}e \right) - \frac{1}{N_2} \cdot \sum_{n_{2}=1}^{N_{2}} \log_2 \left[ \sum_{m_{2}=1}^{N_{2}} \frac{1} {2 + \rho \alpha_2^2 b_{2,2}^{(n_2m_2)2} } \right].
    \end{aligned}
	\end{array}
    \label{eq:I_LB}
    \end{equation}
\hrulefill
\end{figure*}

\normalsize

\section{Simulation Results}
In this section, we provide serveral simulation results to confirm our proposed lower bound and asymptotic analysis. Besides, in order to clarify the benefits of the proposed SM-NOMA system, we provide the MISO-NOMA scheme and the time division multiple access method based on SM (SM-TDMA) as the counterparts.

In simulations for the SM system, we set $M = 4$ BS antennas, $K = 2$ users by user pairing strategies, and each user uniform-randomly selects one of the $M$ transmit antennas. Therefore, the number of required RF chains is $N_{\text{RF}} = 2$. Moreover, we assume that the receiver side has no channel state information (CSI) feedback, so the zero-mean complex Gaussian distributed channels are considered without precoding. In order to obtain the exact value of the targeted MI of the SM system, we adopt the Monte Carlo method to calculate the entropy $h(A)$, i.e., $h(A) = -E_{a \sim f_A(a)} \left\{ \log_2 \left[ f_A(a) \right] \right\}$. Besides, in MISO-NOMA scheme, considering the fairness of comparison, we set the number of TAs as $M' = 2$ without precoding, which also needs $N'_{\text{RF}} = 2$ RF chains.

\begin{figure}
\center{\includegraphics[width=1\linewidth]{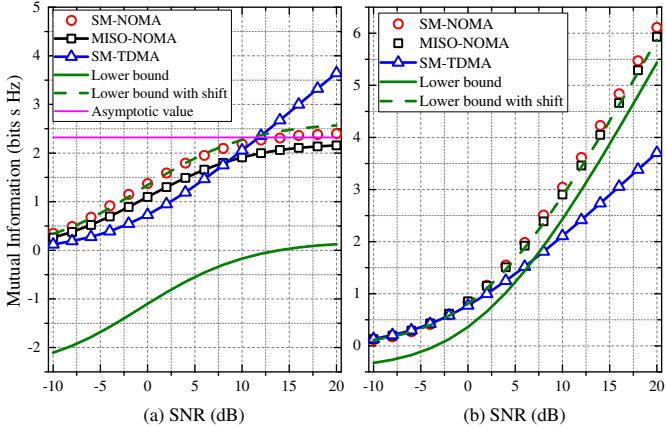}}
\caption{MI of different systems and its lower bound of the SM-NOMA system, given $\alpha_1^2 = 4$ and $\alpha_2^2 = 1$: (a) $I \left( Y_{1,1}; X, B_{1,1} \right)$; (b) $I \left( Y_{2,2}; X, B_{2,2} \right)$.}
\label{fig:Results1}
\end{figure}

\begin{figure}
\center{\includegraphics[width=1\linewidth]{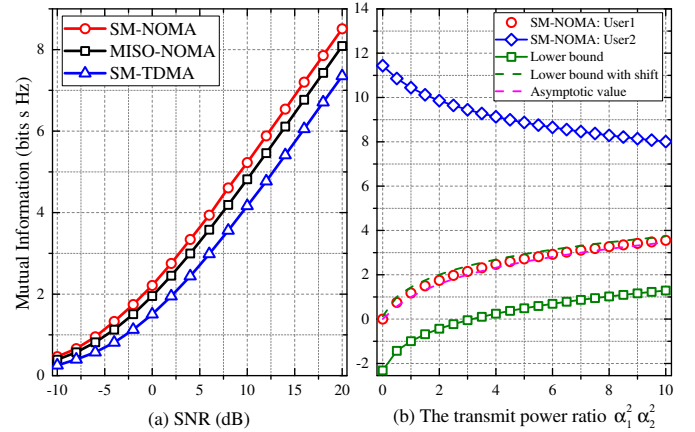}}
\caption{(a) Sum MI of different systems, given $\alpha_1^2 + \alpha_2^2 = 5$; (b) $I \left( Y_{2,2}; X, B_{2,2} \right)$ and $I \left( Y_{1,1}; X, B_{1,1} \right)$ at SNR $=30$ dB.}
\label{fig:Results2}
\end{figure}

Fig.\ref{fig:Results1} shows $I \left( Y_{1,1}; X, B_{1,1} \right)$ and $I \left( Y_{2,2}; X, B_{2,2} \right)$ of different systems as well as the lower bound\footnote{According to \eqref{eq:I_LB}, $I \left( Y_{1,1}; X, B_{1,1} \right)$ and $I \left( Y_{2,1}; X, B_{2,1} \right)$ has almost the same property and lower bound, so we only illustrate the former.} in SM-NOMA system. Obviously, our proposed SM-NOMA system outperforms the conventional MISO-NOMA system. Compared to the SM-TDMA system, although the MI performance of the first user is worse at high SNR regions, the proposed SM-NOMA system still has a larger sum MI, as shown in Fig.\ref{fig:Results2} (a).

In addition, our proposed lower bounds are confirmed with the aforementioned constant shift $C_{1,1}$ and $C_{2,2}$. Because of $h(A)$'s relatively loose upper bound, $I_{\text{LB}} \left( Y_{1,1}; X, B_{1,1} \right)$ has much weaker bound tightness than $I_{\text{LB}} \left( Y_{2,2}; X, B_{2,2} \right)$ and $C_{1,1} > C_{2,2}$. Meanwhile, because of the SIC decoding order, $I \left( Y_{2,2}; X, B_{2,2} \right) > I \left( Y_{1,1}; X, B_{1,1} \right)$, which is shown in Fig.\ref{fig:Results2} (b). As SNR increases, $I \left( Y_{1,1}; X, B_{1,1} \right)$ changes from power-limited to interference-limited and approaches a fixed value in high SNR region, which is analyzed in \eqref{eq:HighSNR}. The only way to enhance $I \left( Y_{1,1}; X, B_{1,1} \right)$ in high SNR region is to increase the transmit power ratio $\frac{\alpha_1^2}{\alpha_2^2}$, as shown in Fig.\ref{fig:Results2} (b).

\section{Conclusion}
In this paper, we propose and analyze a novel SM-NOMA system from the point of view of its MI. The SE performance of our proposed SM-NOMA system is confirmed by simulation results. In our future work, we will analyze the bit error ratio (BER) of the proposed system and focus on the optimization of transmitting power allocation and extend our proposed SM-NOMA system to the generalized SM (GSM) scenarioes.


\end{document}